\documentclass[aps,prl,showpacs,twocolumn]{revtex4}

\usepackage{amsmath}
\usepackage{graphicx}

\begin{document}

\title{Internal Friction and Vulnerability of Mixed Alkali Glasses}

\author{Robby Peibst}
\author{Stephan Schott}
\author{Philipp Maass} 
\affiliation{Institut f\"ur Physik, Technische Universit\"at Ilmenau,
  98684 Ilmenau, Germany}
\email{Philipp.Maass@tu-ilmenau.de}
\homepage{http://www.tu-ilmenau.de/theophys2}

\date{February 24, 2005}

\begin{abstract}
  
  Based on a hopping model we show how the mixed alkali effect in
  glasses can be understood if only a small fraction $c_{\rm V}$ of
  the available sites for the mobile ions is vacant. In particular, we
  reproduce the peculiar behavior of the internal friction and the
  steep fall (``vulnerability'') of the mobility of the majority ion
  upon small replacements by the minority ion. The single and mixed
  alkali internal friction peaks are caused by ion-vacancy and ion-ion
  exchange processes. If $c_{\rm V}$ is small, they can become
  comparable in height even at small mixing ratios.  The large
  vulnerability is explained by a trapping of vacancies induced by the
  minority ions.  Reasonable choices of model parameters yield typical
  behaviors found in experiments.

\end{abstract}

%66.30.Dn Theory of diffusion and ionic conduction in solids
%66.30.Hs Self-diffusion and ionic conduction in nonmetals
%82.45.Gj Electrolytes (for polyelectrolytes, see also 82.35.Rs and 82.45.Wx;
%         see also 66.30.Hs Self-diffusion and ionic conduction in nonmetals)

\maketitle

The mixed alkali effect (MAE) is a key problem for understanding ion
transport processes in glasses and refers to strong changes in
transport properties upon mixing of two types of mobile ions (for
reviews, see \cite{Day:1976,Ingram:1994}). Fundamental for the MAE is
the behavior of the tracer diffusion coefficients $D_{\rm A}$ and
$D_{\rm B}$ of two types of ions A and B. With increasing number
fraction $x$ of B ions, i.e.\ with successive replacement of A by B
ions, $D_{\rm A}$ decreases while $D_{\rm B}$ increases. These changes
are reflected in the activation energies $E_{\rm A,B}(x)\sim-k_{\rm
  B}T\log D_{\rm A,B}(x)$, so that at low temperatures $D_{\rm
  A,B}(x)$ vary by many orders of magnitude. As a consequence, the
ionic conductivity $\sigma(x)\sim (1-x)D_{\rm A}(x)+xD_{\rm B}(x)$
runs through a deep minimum close to the intersection point of $D_{\rm
  A}$ and $D_{\rm B}$.

Much progress was made in the past to explain the MAE
\cite{Maass/etal:1992,Hunt:1994,Greaves/Ngai:1995,Tomozawa:1998,Schulz/etal:1999,Maass:1999,Kirchheim:2000,Swenson/etal:2001,Swenson/Adams:2003,Bunde/etal:2004}.
EXAFS \cite{Greaves/etal:1991} and NMR measurements
\cite{Gee/etal:1997}, and in particular recent analyses of neutron and
x-ray scattering data with the reverse Monte Carlo technique
\cite{Swenson/etal:2001} support the picture that a structural
mismatch effect leads to distinct preferential diffusion pathways for
each type of mobile ion. The connectivity and interference of these
pathways determines the long-range ion mobilities. On this theoretical
basis alone, however, main properties of the MAE are still not well
understood.

One property is the large ``vulnerability'', that means the steep
decrease of the diffusion coefficient of the majority ion, e.g.\ A,
with beginning replacement by the minority ion B
\cite{Ingram/Roling:2002,Moynihan/Lesikar:1981}. In many mixed alkali
glasses, $\ln D_{\rm A}(x)$ is a convex function for small $x$,
reflecting that $E_{\rm A}(x)$ is a concave function in these systems
\cite{McVay/Day:1970,Voss/etal:2004}. Why is the increase of $E_{\rm
  A}$ largest at beginning replacement, where the influence of the
minority ion should be weak?

Another quantity poorly understood so far is the internal friction
(for a review, see \cite{Zdaniewski/etal:1979}). Measurements on mixed
alkali glasses show that the ionic motion leads to two mechanical loss
peaks, a single alkali peak (SP) and a mixed alkali peak (MP). The
peak frequency $\tau_{\rm s}^{-1}$ of the SP has (nearly) the same
activation energy $E_{\rm s}$ as the diffusivity of the more mobile
ion, while the lower peak frequency $\tau_{\rm m}^{-1}$ of the MP
exhibits an activation energy $E_{\rm m}$ that is not simply related
to that of the mobility of the less mobile ion. Particularly puzzling
is that the height $H_{\rm m}$ of the MP can be comparable to the
height $H_{\rm s}$ of the SP even at very small $x$, and that $H_{\rm
  m}$ becomes larger the more similar the two types of ions are
\cite{Day:1972}.

In this Letter we show that these peculiar behaviors can be explained,
if the fraction $c_{\rm V}$ of vacant sites for the mobile ions is
small. Small values $c_{\rm V}=5-10\%$ have been found recently in
molecular dynamics simulations
\cite{Lammert/etal:2003,Habasaki/Hiwatari:2004}, and should be
expected also on general grounds, since empty sites correspond to
local structural configuration of high energy \cite{Dyre:2003}. Here
we argue that the experimental results for the internal friction
provide independent support for this feature.

To reason that conjecture, we note (see eq.~(\ref{eq:q}) below) that
the SP and MP can be attributed to processes, where the more mobile
ions (e.g.\ A) exchange sites with vacancies (AV exchange), and where
the A ions exchange sites with B ions (AB exchange). When denoting by
$c_{\rm A,B}$ the fractions of sites occupied by A,B ions, a
mean-field argument then predicts that $H_{\rm s}\propto c_{\rm
  A}c_{\rm V}$ and $H_{\rm m}\propto c_{\rm A}c_{\rm B}$ should become
comparable if $c_{\rm V}\simeq c_{\rm B}$. Therefore, if it is
possible to find $H_{\rm m}\simeq H_{\rm s}$ at small $x$, where
$c_{\rm B}\sim x$ is small, $c_{\rm V}$ must be small also. Indeed, we
will show in the following that this rough argument can be
substantiated.

First we are confronted with the general problem whether the mismatch
concept is sustainable as mechanism for the MAE if $c_{\rm V}$ is
small. To this end we investigate a lattice model, where the rate
$w_{ij}^\alpha$ for a jump of an $\alpha$ ion ($\alpha={\rm A,B}$)
from site $i$ to a nearest neighbor site $j$ is determined by the site
and barrier energies
\begin{equation}
\epsilon_i^\alpha=-\epsilon_{\rm mis}^\alpha\,\mu_i^\alpha\,,\hspace{2em}
u_{ij}^{\alpha}=u_0^\alpha+
\frac{u_{\rm mis}^\alpha}{2}(\mu_i^\beta+\mu_j^\beta)\,.
\label{eq:energies}
\end{equation}
The mismatch effect is taken into account via the mismatch energies
$\epsilon_i^\alpha>0$, $u_{\rm mis}^{\alpha}>0$, and the structural
variables $\mu_i^\alpha$, $0\le\mu_i^\alpha\le1$ that specify the
$\alpha$-character of site $i$. A site $i$ with large $\mu_i^\alpha$
has a local environment favorably accommodated to an $\alpha$ ion and
hence a low energy $\epsilon_i^\alpha$.  Large values of $\mu_i^{\rm
  A}$ imply a small value of $\mu_i^{\rm B}$ and vice versa. This is
accounted for by taking opposing mean values $\bar\mu_i^{\rm
  A}=1-\bar\mu_i^{\rm B}$ in the distributions of the $\mu_i^{\alpha}$
(see below). Below a threshold value $\mu_c^\alpha$, sites loose their
identity and we require the energy to saturate
($\epsilon_i^\alpha=-\epsilon_{\rm mis}\mu_c^\alpha$ for
$\mu_i^{\alpha}\le\mu_c^\alpha$).

The barriers $u_{ij}^{\alpha}$ in eq.~(\ref{eq:energies}) contain a
bare structural barrier $u_0^\alpha$ (characterizing the activation
energy in the pure systems) and a mismatch barrier that becomes higher
with increasing ``foreign'' $\beta(\ne\alpha)$-characters of the
initial and target sites involved in a jump of an $\alpha$ ion. This
is necessary in order to obtain low mobilities for the minority ion in
the dilute regime \cite{entropy-comm}. Within an Anderson-Stuart like
picture \cite{Anderson/Stuart:1954} it can be physically understood by
the fact that larger minority ions have to bring up an additional
elastic strain energy when open up doorways, and that smaller minority
ions have to surmount a higher saddle point energy associated with the
Coulomb potential of the counter ions.

The disorder in the glass is reflected in the distributions of the
$\mu_i^\alpha$. Due to the local accommodation of the network
structure to the ions in course of the freezing process, the numbers
of sites with large $\mu_i^{\rm A}$ and $\mu_i^{\rm B}$ scale with
$(1-x)$ and $x$, respectively. In addition we take into account
short-range correlation between the $\mu_i^\alpha$. A convenient
(technical) way to generate corresponding distributions (the detailed
form is not important here) is as follows: We draw $\mu_i^\alpha$ from
truncated Gaussians
($p(\mu_i^\alpha)\propto\exp[-(\mu_i^\alpha-\bar\mu_i^\alpha)^2
/2\Delta_\alpha^2]$ for $\mu_i^\alpha\in[0,1]$) with $\bar\mu_i^{\rm
  A}=\sum_j\eta_j/(z+1)=1-\bar\mu_i^{\rm B}$, where the sum runs over
site $i$ and its nearest neighbors, and $\eta_j$ are random binary
numbers equal to one with probability $(1-x)$.

Kinetic Monte Carlo simulations are performed in a simple cubic
lattice with periodic boundary conditions, using Metropolis hopping
rates $w_{ij}^\alpha=\nu_\alpha\exp(-u_{ij}^\alpha/k_{\rm B}T)
\min\bigl(1,\exp[-(\epsilon_j^\alpha-\epsilon_i^\alpha)/k_{\rm
  B}T]\bigr)$ ($\nu_\alpha$ are attempt frequencies).  We choose
$c_{\rm V}=0.05$ and consider, for computational convenience, a
symmetric set of parameters $\nu_{\rm A}=\nu_{\rm B}\equiv\nu$,
$\epsilon_{\rm mis}^{\rm A}=\epsilon_{\rm mis}^{\rm B}\equiv\epsilon_{\rm
  mis}$, $\ldots$ To determine the tracer diffusion coefficients and
internal friction as functions of $x$ and reduced temperature
$\theta\equiv k_{\rm B}T/\epsilon_{\rm mis}$, we fix the remaining
parameters to $\mu_c=2/7$, $\Delta=1/14$, and $u_{\rm
  mis}=0.7\epsilon_{\rm mis}$. The parameter $\epsilon_{\rm mis}$
characterizes the difference between the two types of mobile ions
(e.g.\ with respect to size).

%****************************************************************
\begin{figure}[t!]
\centering
\includegraphics[width=0.45\textwidth]{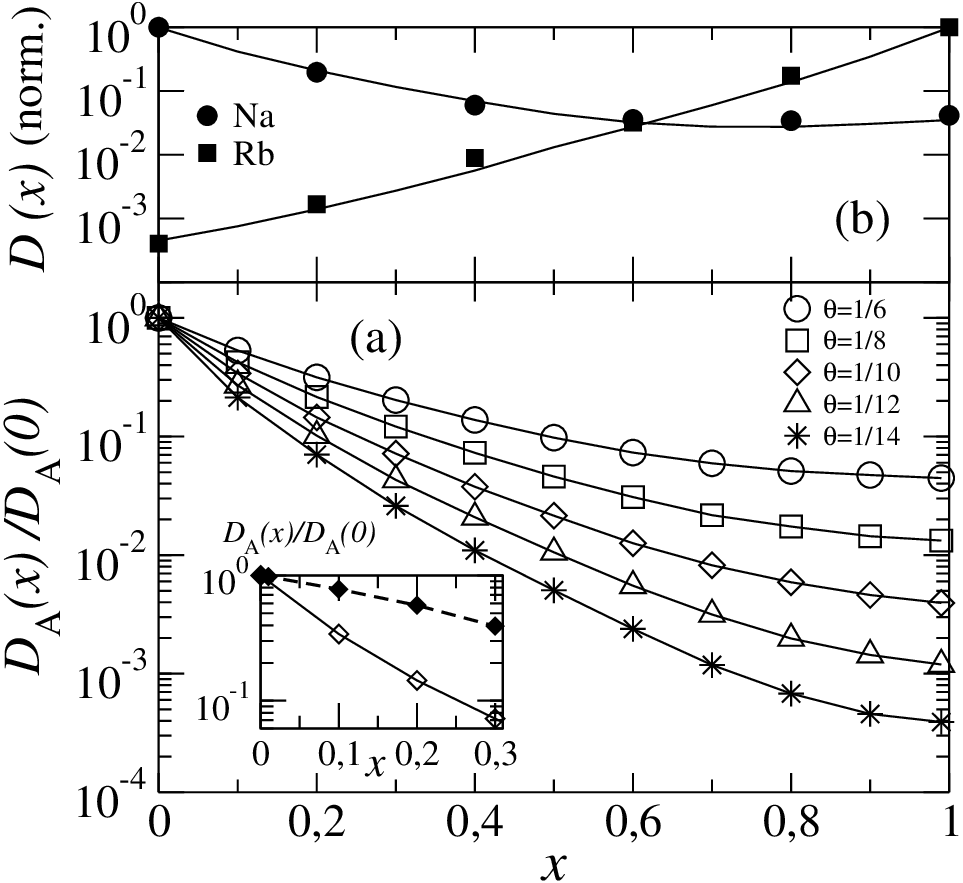}
\caption{{\it (a)} Normalized diffusion coefficient $D_{\rm A}(x)/D_{\rm
    A}(0)$ of A ions for 5 different reduced temperatures
  $\theta=k_{\rm B}T/\epsilon_{\rm mis}$. The inset shows $D_{\rm
    A}(x)$ for small $x$ at $\theta=1/10$ (open symbols) in comparison
  with the results in the absence of a correlation induced trapping
  effect (full symbols).  {\it (b)} Measured data (symbols) for
  $(1-x)$Na$_2$O-$x$Rb$_2$-4B$_2$O$_3$ glasses at 652K (redrawn from
  \cite{Voss/etal:2004}) and fit by the model (lines) with
  $\theta=1/9$ and asymmetric mismatch barriers $u^{\rm B}_{\rm
    mis}=2u^{\rm A}_{\rm mis}=\epsilon_{\rm mis}$.}
\label{fig:diffusion}
\vspace*{-0.5cm}
\end{figure}
%****************************************************************

Figure~\ref{fig:diffusion}a shows the normalized diffusion coefficient
$D_{\rm A}(x)/D_{\rm A}(0)$ as a function of $x$ for 5 different
$\theta$ [due to the symmetric parameters we do not plot $D_{\rm
  B}(x)=D_{\rm A}(1-x)$]. The curves exhibit the typical behavior
found in experiment: $D_{\rm A}(x)$ decreases by many orders of
magnitude when A ions are replaced by B ions. This effect becomes
stronger with lower temperature or stronger difference $\epsilon_{\rm
  mis}$ between the two types of mobile ions, i.e.\ with larger
$\theta=k_{\rm B}T/\epsilon_{\rm mis}$.  We thus conclude that the
mismatch concept is sustainable for small $c_{\rm V}$.

In particular, $\log[D_{\rm A}(x)/D_{\rm A}(0)]=-E_{\rm A}(x)/k_{\rm
  B}T$, exhibits a convex curvature for small $x$, and $E_{\rm A}(x)$
a concave curvature (see Fig.~\ref{fig:eact}a; note that one must add
$u_0$ to obtain the total activation energy $E_{\rm A}^{\rm
  tot}(x)=E_{\rm A}(x)+u_0$). The large vulnerability is caused by a
trapping effect: For small $x$, there exist regions in the glass where
the network structure tended to accommodate to the ``foreign'' B ions.
Because of the spatial correlations, the regions consist of several
sites with smaller A character $\mu_i^{\rm A}$ than on average. These
sites are not only favorably occupied by B ions, but also favorably
occupied by vacancies, because this allows more of the majority A ions
to occupy well accommodated sites with large $\mu_i^{\rm A}$ close to
one. Vacancies hence become trapped in regions more favourably
accomodated to the the foreign B ions and they have to bring up an
additional activation energy to promote the diffusion of A ions. The
corresponding reduction of $D_{\rm A}(x)$ is strongest for $x\to0$,
since with increasing $x$ the number of sites with values $\mu_i^{\rm
  A}$ close to one decreases. The importance of the trapping effect
\cite{trap-comm} is demonstrated in the inset of
Fig.~\ref{fig:diffusion}a, where $D_{\rm A}(x)$ is shown in comparison
with the curve obtained in the absence of spatial correlations
($\bar\mu_i^{\rm A}=\eta_i^{\rm A}$ instead of $\bar\mu_i^{\rm
  A}=\sum_j\eta_j/(z+1)$) but otherwise the same parameters).

Fig.~\ref{fig:diffusion}a resembles well the experimental behavior. By
choosing asymmetric parameters one can reproduce measured curves as
shown in Fig.~\ref{fig:diffusion}b with reasonable parameter values
(if we require $\theta=1/9$ in Fig.~\ref{fig:diffusion}b to represent
  the experimental temperature 652K, we would obtain a mismatch energy
  $\epsilon_{\rm mis}=0.5$eV). A detailed fitting to experimental
  results, however, is not our aim here (and has limitations due the
  simplicity of the model).

%****************************************************************
\begin{figure}[t!]
\includegraphics[width=0.45\textwidth]{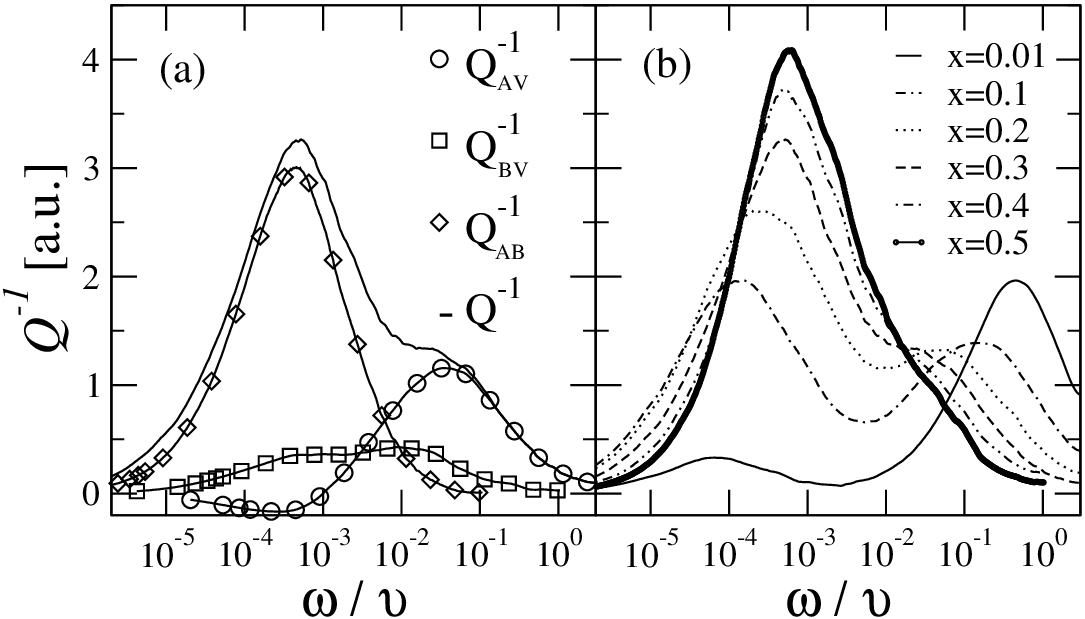}
\caption{{\it (a)} Internal Friction $Q^{-1}$ and spectral components
  (cf.\ eq.~\ref{eq:q}) as a function of $\omega$ for $x=0.3$ and
  $\theta=1/10$. {\it (b)} Change of the internal friction $Q^{-1}$
  with $x$ for $\theta=1/10$.}
\label{fig:intfric}
\vspace*{-0.2cm}
\end{figure}
%****************************************************************

Rather, we next explore how the behavior in the internal friction
$Q^{-1}(\omega,T)$ can be understood. To this end, we consider a small
modulation $\Delta \epsilon_i^\alpha(t)=a_0\,\Re[\chi_i^\alpha
e^{-i\omega t}]$ of the site energies in response to an oscillatory
shear stress with amplitude $a_0$ and frequency $\omega$
\cite{Knoedler/etal:1995}. The complex coupling constants
$\chi_i^\alpha$ fluctuate due to the disorder in the glass. Without
making specific assumptions regarding their correlation properties, we
employ a random phase ansatz and regard them to be uncorrelated for
different sites $i$ and different types $\alpha$,
$[\chi_i^\alpha\chi_j^{\beta\,\star}]_{\rm
  av}\propto\delta_{\alpha\beta}\delta_{ij}$, and uncorrelated with
the site energies of the unperturbed system
$[\chi_i^\alpha\epsilon_i^\alpha]_{\rm av}=0$ ($[\ldots]_{\rm av}$
denotes a disorder average). Introducing the occupation numbers
$n_i^\alpha(t)$, i.e.\ $n_i^\alpha(t)=1$ if site $i$ is occupied by
species $\alpha$ ($\alpha={\rm A,B,V}$ here) at time $t$ and zero
else, linear response theory then expresses the internal friction in
terms of the Fourier cosine transforms of the exchange correlation
functions $\langle n_i^\alpha(t)n_j^\beta(0)\rangle$
($\langle\ldots\rangle$ denotes a thermal average). The result is
\begin{eqnarray}
Q^{-1}(\omega,T)&=&\frac{\kappa\,\omega}{k_{\rm B}T}\Bigl[
\sum_{\alpha={\rm A,B}}
c_{\alpha{\rm V}}S_{\alpha{\rm V}}+2c_{\rm AB}S_{\rm AB}\Bigr]\label{eq:q}\\
S_{\alpha\beta}(\omega,T)&=&\int_0^\infty
  dt\,C_{\alpha\beta}(t)\cos(\omega t)\,,\label{eq:sab}
\end{eqnarray}
where $\kappa$ is constant \cite{kappa-comm},
$c_{\alpha\beta}\!=\![\langle n_i^\alpha\rangle\langle 
n_i^\beta\rangle]_{\rm av}$,
and $C_{\alpha\beta}(t)\!=$ $1-\![\langle
n_i^\alpha(t)n_i^\beta\rangle]_{\rm av}/c_{\alpha\beta}$ are the
normalized exchange correlation functions [$C_{\alpha\beta}(0)=1$ and
$C_{\alpha\beta}(t\to\infty)=0$].

Figure~\ref{fig:intfric}a shows $Q^{-1}$ as a function of $\omega$ for
$x=0.3$ and $\theta=1/10$, together with the spectral components
$Q^{-1}_{\alpha\beta}\propto(\omega/\theta)c_{\alpha\beta}S_{\alpha\beta}$
in eq.~(\ref{eq:q}). Two arise from exchange processes of the ions
with the vacancies (AV and BV exchange) and one arises from exchange
processes of the two types of ions (AB exchange). The exchange of
vacancies with the majority ions (AV) leads to the SP and the AB
exchange to the MP, while the BV exchange yields a broad spectral
contribution that is masked by the SP and MP, and cannot be resolved in
$Q^{-1}$. The overall behavior of $Q^{-1}$ with varying $x$
for $\theta=1/10$ is shown in Fig.~\ref{fig:intfric}b. Due to the
small $c_{\rm V}$, the MP can be well identified already for $x=0.01$,
in agreement with the argument given above. With increasing $x$ the MP
becomes higher and moves towards larger frequencies, while the SP
becomes lower and moves towards smaller frequencies. At $x=0.5$ the SP
cannot be resolved any longer.

\begin{figure}[t!]
\includegraphics[width=0.45\textwidth]{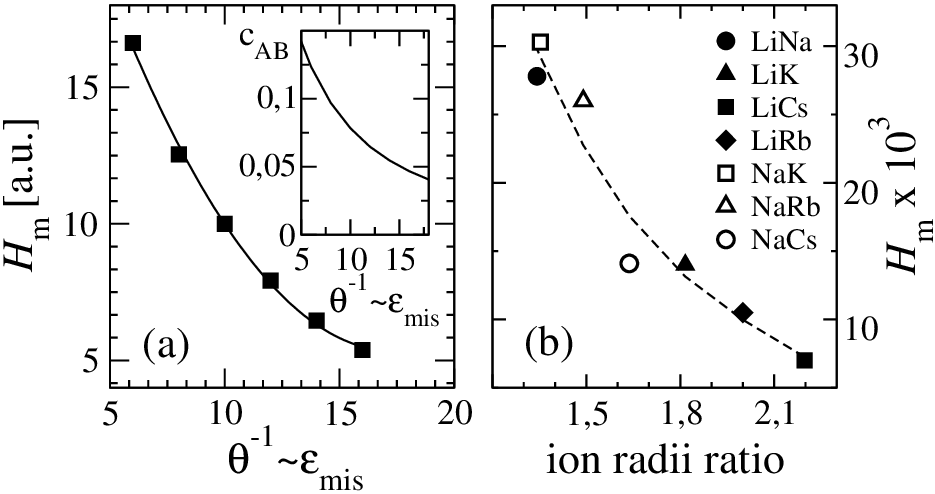}
\caption{{\it (a)} $H_{\rm m}$ in
  dependence of $\theta^{-1}=\epsilon_{\rm mis}/k_{\rm B}T$ at
  $x=0.5$. The inset shows the decrease of $c_{\rm AB}$ with
  increasing $\theta^{-1}\propto\epsilon_{\rm mis}$ (from analytical
  calculation).  {\it (b)} $H_{\rm m}$ as a function of ion radii
  ratio of various mixed alkali pairs for $x=0.5$ and
  $\omega/2\pi=0.4$Hz fixed.}
\label{fig:height-mp}
\vspace*{-0.2cm}
\end{figure}

Remarkably, also the puzzling rise of the height $H_{\rm m}$ with
increasing similarity of the two types of mobile ions (i.e.\ decreasing
$\epsilon_{\rm mis}$) is reproduced, see Fig.~\ref{fig:height-mp}a.
$H_{\rm m}$ at $x=0.5$ becomes higher with decreasing $\epsilon_{\rm
  mis}$ (we regard $T$ to be fixed here). The reason for this behavior
lies in the weighting of the AB peak by the disorder averaged product
$c_{\rm AB}\!=\![\langle n_i^{\rm A}\rangle \langle n_i^{\rm
  B}\rangle]_{\rm av}$ of the equilibrium occupations $\langle
n_i^{\rm A}\rangle$, $\langle n_i^{\rm B}\rangle$ in eq.~(\ref{eq:q}).
With increasing mismatch $\epsilon_{\rm mis}$, A and B ions share the
same sites with lower probability (see inset of
Fig.~\ref{fig:height-mp}a) and accordingly $H_{\rm m}$ decreases. The
behavior is reminiscent to the variation of $H_{\rm m}$ with the
fraction of ion radii observed in experiments at $x=0.5$ for a fixed
frequency, see Fig.~\ref{fig:height-mp}b
\cite{height-mp-comm}.

When denoting by $E_{\alpha\beta}$ the activation energies of the
$Q^{-1}_{\alpha\beta}$ peak frequencies $\tau_{\alpha\beta}^{-1}$, we
can define by $E_{\rm s}=\min(E_{\rm AV},E_{\rm BV})$ and $E_{\rm
  m}=E_{\rm AB}$ the activation energies of the SP and MP. The
activation energies $E_{\alpha\beta}$ obtained from Arrhenius plots
(for the same $\theta$ values as in Fig.~\ref{fig:diffusion}) are
shown in Fig.~\ref{fig:eact}a together with the diffusion activation
energies $E_<\equiv\min(E_{\rm A},E_{\rm B})$ and
$E_>\equiv\max(E_{\rm A},E_{\rm B})$ of the more and less mobile ion.
We find that $E_{\rm s}$ ($=\min(E_{\rm AV},E_{\rm BV})$) follows
closely $E_<$, in agreement with measurements
\cite{Zdaniewski/etal:1979}.  By contrast, $E_{\rm m}$ ($=E_{\rm AB}$)
differs from $E_>$. The deviations found in Fig.~\ref{fig:eact}a are
similar to those found in experiments, see Fig.~\ref{fig:eact}b
\cite{esmall-comm}.

%****************************************************************
%****************************************************************

To show $Q^{-1}$ as function of temperature for fixed frequency, as
typically obtained in experiments, we use time-temperature scaling to
transform $Q^{-1}$ from frequency to temperature space. Taking the
$\theta$ values from Fig.~\ref{fig:diffusion}a, we first checked, that
a scaling $\omega
S_{\alpha\beta}(\omega,T)=F_{\alpha\beta}(\omega\tau_{\alpha\beta}(T))$
is approximately valid. By applying this scaling we then replot
$Q^{-1}$ from Fig.~\ref{fig:intfric}b as function of $T$ in
Fig.~\ref{fig:eact}c for small $x$ and parameter values $\epsilon_{\rm
  mis}=1$eV, $u_0=0.5$eV, and $\omega/\nu=10^{-12}$. These values
yield activation energies $E_{\rm A}^{\rm tot}(x)=E_{\rm
  A}+u_0\simeq0.6\ldots1.2$eV comparable to those found in
experiments, cf.\ Fig.~\ref{fig:eact}b. The choice
$\omega/\nu=10^{-12}$ corresponds to $\omega\simeq1$Hz, when
$\nu\simeq10^{12}$Hz is a typical attempt frequency. Using this rough
parameter estimate, peak temperatures in Fig.~\ref{fig:eact}c are
obtained that reflect the experimental scenario in
Fig.~\ref{fig:eact}d [note that the glass transition and the
``non-bridgening oxygen'' (nbO) peak \cite{Zdaniewski/etal:1979} are
not taken into account in the present approach].

In summary, we have shown how the mixed alkali effect in glasses,
including the vulnerability and the peculiar features of the internal
friction, can be understood based on the mismatch effect in the
presence of a small vacancy concentration. Due to the small $c_{\rm
  V}$, $H_{\rm m}$ can become comparable to $H_{\rm s}$ even at small
concentrations of the minority ion, and $H_{\rm m}$ increases, when it
becomes easier for both types of ions to share the same sites. The
large vulnerability can be connected to a trapping of vacancies
induced by the minority ions. Reasonable choices of parameters allowed
us to faithfully reproduce typical behaviors found in experiments.

%****************************************************************
\begin{figure}[t!]
\includegraphics[width=0.5\textwidth]{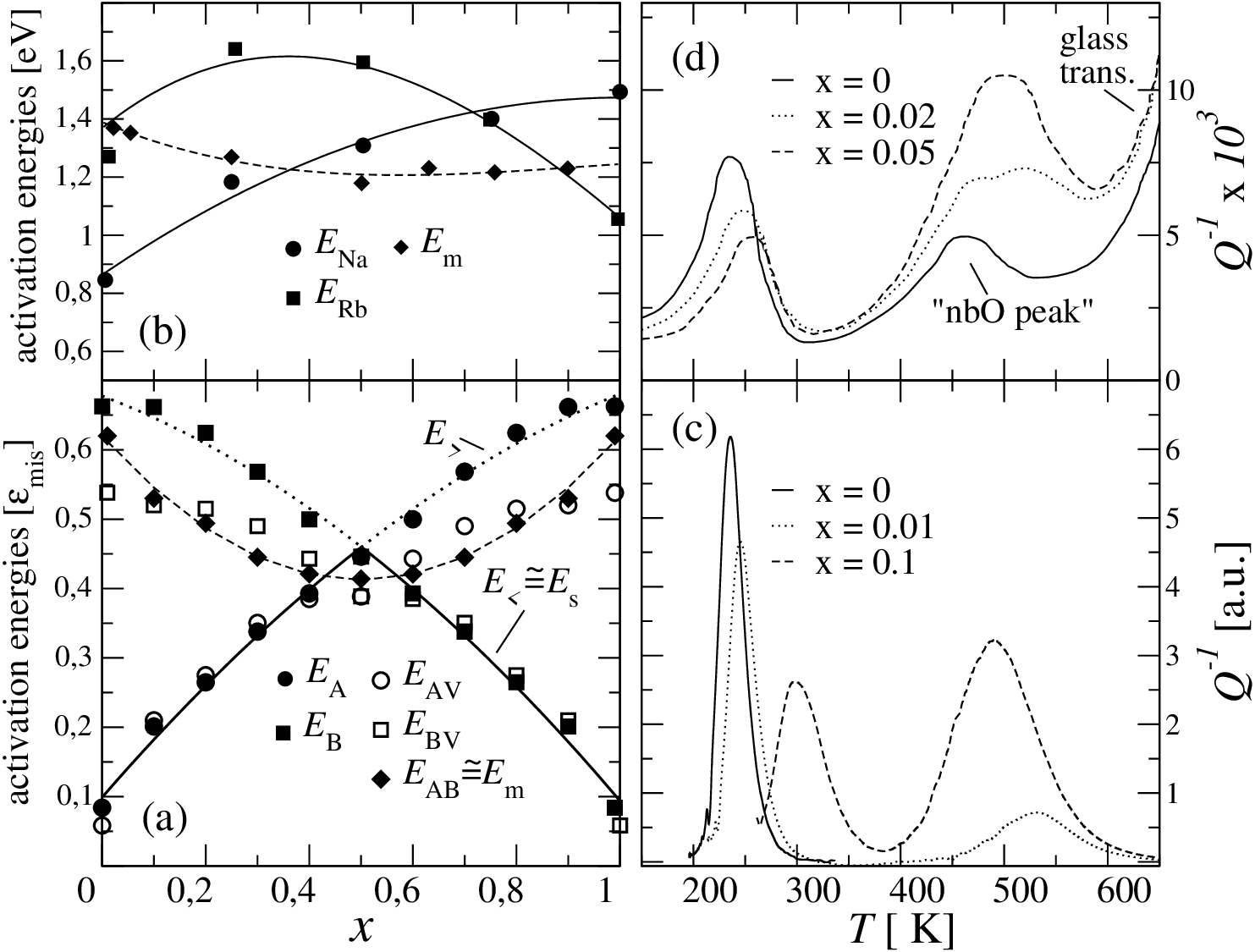}
\caption{{\it (a)} Activation energies of the tracer diffusion
  coefficients shown in Fig.~\ref{fig:diffusion}, and of the spectral
  components of the internal friction. {\it (b)} $E_{\rm Na}$, $E_{\rm
    Rb}$, and $E_{\rm m}$ in $(1-x)$Na$_2$O-$x$Rb$_2$O-3SiO$_2$
  glasses (redrawn from \cite{McVay/Day:1970}). {\it (c)} $Q^{-1}$
  calculated from Fig.~\ref{fig:intfric} after applying
  time-temperature scaling for $\epsilon_{\rm mis}=2u_0=1$eV and
  $\omega/\nu=10^{12}$ (see text).  {\it (d)} $Q^{-1}$ in
  $(1-x)$Na$_2$O-$x$Rb$_2$O-3SiO$_2$ glasses for $\omega/2\pi=0.4$Hz
  (redrawn from \cite{McVay/Day:1970}).}
\label{fig:eact}
\vspace*{-0.3cm}
\end{figure}
%****************************************************************

We thank W.~Dieterich, J.~Habasaki, and M.~D.~Ingram for valuable
discussions.

\enlargethispage{0.1cm}


\begin{thebibliography}{99}
\bibitem{entropy-comm} When a mismatch in the energy barriers is
  absent, the minority ions would, in the dilute limit, be driven away
  from their favorable low-energy sites by entropic forces and assume
  mobilities comparable to the majority ions.
\bibitem{kappa-comm} For the coefficient $\kappa$ one finds
  $\kappa\!=\!2\pi [|\chi_i^\alpha|^2]_{\rm av}/M'a^3$, where $M'$ is
  the real part of the shear modulus, which is governed by the
  (elastic) network response and thus very weakly dependent on
  frequency ($a$: mean jump distance).
\bibitem{height-mp-comm} When the ion ratio approaches one (as for the
  K/Rb pair), the correlation to $\epsilon_{\rm mis}$ should become
  weaker and other properties (e.g.\ polarizabilities) will dominate
  $\epsilon_{\rm mis}$.
\bibitem{esmall-comm} It is interesting to note that the activation
  energy corresponding to the exchange of vacancies with the less
  mobile ions does not follow the diffusion activation energy of the
  less mobile ion.
\end{thebibliography}
\end{document}